# Broad Spectral Tuning of Ultra-Low Loss Polaritons in a van der Waals Crystal by Intercalation

Javier Taboada-Gutiérrez[1,2], Gonzalo Álvarez-Pérez[1,2], Jiahua Duan[1,2], Weiliang Ma[3], Kyle Crowley[4], Iván Prieto[5], Andrei Bylinkin[6,7], Marta Autore[7], Halyna Volkova[8], Kenta Kimura[9], Tsuyoshi Kimura[9], M.-H. Berger[8], Shaojuan Li[10], Qiaoliang Bao[3], Xuan P.A. Gao[4], Ion Errea[6,11,12], Alexey Nikitin[6,13], Rainer Hillenbrand[7,13], Javier Martín-Sánchez[1,2*], and Pablo Alonso-González[1,2*]

1. Departamento de Física, Universidad de Oviedo, Oviedo, Spain.
2. Nanomaterials and Nanotechnology Research Center (CINN), El Entrego, Spain.
3. Department of Materials Science and Engineering, and ARC Centre of Excellence in Future Low-Energy Electronics Technologies (FLEET), Monash University, Clayton, Victoria, Australia.
4. Department of Physics, Case Western Reserve University, Cleveland, USA.
5. Institute of Science and Technology (IST), Viena, Austria.
6. Donostia International Physics Center (DIPC), Donostia/San Sebastián, Spain.
7. CIC nanoGUNE, Donostia/San Sebastián, Spain.
8. Centre des Matériaux, CNRS UMR 7633 – PSL Research University, MINES ParisTech, 91003 Evry Cedex, France
9. Department of Advanced Materials Science, University of Tokyo, Kashiwa, Chiba 277-8561, Japan
10. Institute of Functional Nano and Soft Material(FUNSOM), Soochow University, Suzhou 215123, China
11. Fisika Aplikatua 1 Saila, University of the Basque Country (UPV/EHU), Donostia/San Sebastián, Spain.
12. Centro de Física de Materiales (CSIC-UPV/EHU), Donostia/San Sebastián, Spain.
13. IKERBASQUE, Basque Foundation for Science, Bilbao, Spain.

*e-mail: pabloalonso@uniovi.es, javiermartin@uniovi.es

## ABSTRACT

**Phonon polaritons (PhPs) – light coupled to lattice vibrations – in polar van der Waals (vdW) crystals are promising candidates for controlling the flow of energy at the nanoscale due to their strong field confinement, anisotropic propagation and ultra-long lifetime in the picosecond range[1-5]. However, the lack of tunability of their narrow and material-specific spectral range – the Reststrahlen Band (RB) – severely limits their technological implementation. Here, we demonstrate that intercalation of Na atoms in the vdW semiconductor α-$V_2O_5$ enables a broad spectral shift of RBs, and that the PhPs excited show ultra-low losses (lifetime of 4 ± 1 ps), similar to PhPs in the non-intercalated crystal (lifetime of 6 ± 1 ps). We expect our intercalation method to be applicable to other vdW crystals, opening the door for the use of PhPs in broad spectral bands in the mid-infrared domain.**

In recent years, polar van der Waals (vdW) crystals have become ~~ideal~~ excellent platforms to study and manipulate light at the nanoscale[6-9], the goal of the ~~vibrant~~ burgeoning field of nano-optics. Their reduced dimensionality and intrinsic anisotropy have allowed the discovery of infrared-active phonon polaritons (PhPs) – light coupled to lattice vibrations – with extraordinary properties. Prominent examples are PhPs in the dielectric hexagonal boron nitride (h-BN) — which uncover ultra-strong confinement, slow propagation, and out-of-plane hyperbolic behavior, allowing the observation of exotic optical phenomena such as ray propagation or hyper-lensing effects[10] —, and in the polar vdW semiconductor $\alpha$-MoO$_3$ — revealing in-plane anisotropic propagation (hyperbolic or elliptic) with ~~record~~ ultra-low losses[2] (lifetimes of 8 ps) —, offering ~~new~~ opportunities for planar nanotechnologies aiming a directional control of light-matter interactions at the nanoscale.

However, despite the extraordinary properties of these PhPs in vdW crystals, there is still an important technological drawback for their implementation into nanophotonics technologies, namely the narrow and material-dependent spectral band where they exist (the so-called Reststrahlen band (RB), defined between the transverse (TO) and longitudinal optic (LO) phonon frequencies). Although attempts to spectrally tune the RB in bulk polar materials have been recently tackled[11,12], a broad spectral tuning of a RB in a polar vdW material, and consequently of the PhPs supported within it, has remained elusive.

In this work, we introduce intercalation as an efficient route to spectrally tune PhPs in a ~~novel~~ polaritonic vdW crystal, the metal oxide $\alpha$-V$_2$O$_5$. By near-field imaging of Na-intercalated $\alpha$-V$_2$O$_5$ (forming the crystal $\alpha$'-(Na)V$_2$O$_5$) we reveal a RB shift of ~30 cm$^{-1}$ (60% of the initial RB width), and consequently of the PhPs supported within it. Importantly, PhPs in the intercalated $\alpha$'-(Na)V$_2$O$_5$ crystal exhibit ultra-low loss (lifetimes of 4±1 ps) and in-plane anisotropic propagation, similar to PhPs in pristine $\alpha$-V$_2$O$_5$ (lifetimes of 6±1 ps), evidencing that the intercalation of atoms in between the vdW layers does not significantly affect the polaritonic properties of the crystal.

The optical image in Fig. 1a shows the characteristic rectangular shape of $\alpha$-V$_2$O$_5$ thin flakes studied along this work. This asymmetric geometry is inherited from the strong anisotropy of its crystal lattice[13,14], which is illustrated in the diagram of Fig. 1b. As can be seen, the crystalline unit cell of $\alpha$-V$_2$O$_5$ shows an orthorhombic structure where the three inequivalent oxygen positions (denoted O$_{1-3}$) with respect to the vanadium atom give rise to asymmetric V–O bonds along the three different crystalline axes. This strong biaxial anisotropy is also translated into the optical properties of the crystal (Supplementary Information) as observed in Fig. 1c, where the dielectric permittivity of $\alpha$-V$_2$O$_5$[15] along the three crystalline axes is shown. We can see up to three spectral bands in the plotted range (in color) where at least one of the permittivity components is negative, thus indicating the existence of three RBs (denoted RB$_{1-3}$) and the possibility of supporting PhPs within them. Given that only one (as in RB$_1$ and RB$_3$) or two (as in RB$_2$) components of the permittivity are negative, instead of the three components (as in isotropic media), PhPs in $\alpha$-V$_2$O$_5$ are expected to be strongly anisotropic with either elliptical or hyperbolic propagation[2,5,16,17].

To probe the excitation of PhPs in $\alpha$-V$_2$O$_5$, we performed polariton interferometry[18,19] using scattering-type scanning near-field optical microscopy (s-SNOM), which yields nanoscale resolved near-field images together with sample topography (Methods). In Figs. 2a-c we show near-field amplitude images of an $\alpha$-V$_2$O$_5$ flake with thickness $d$ = 105 nm, taken at frequencies belonging to RB$_1$ of $\alpha$-V$_2$O$_5$: $\omega_0$ = 1031 cm$^{-1}$, $\omega_0$ = 1026 cm$^{-1}$ and $\omega_0$ = 1020 cm$^{-1}$, respectively. In the images we observe bright fringes parallel to all flake edges, but with different periodicities depending on the crystal direction, indicating PhPs with in-plane (along the flake) anisotropic propagation[2,5]. This can be better examined by analyzing the profiles shown on the right of the s-SNOM images in Fig. 2b, where the polariton wavelength $\lambda_p$ is about 915 nm along the [100] crystal direction and about 800 nm along the [001] direction (note that in polariton interferometry[18,19] the polaritons excited by the tip propagate away and are back-reflected at the flake edges, giving rise to interference fringes with a spacing $\lambda_p/2$), clearly corroborating an in-plane anisotropic propagation of PhPs. Apart from this extraordinary behavior, the PhPs wavelengths along the [100] and [001] directions also reveal a deep subwavelength-scale polariton confinement $\lambda_p \ll \lambda_0 = 9.75$ μm (where $\lambda_0$ is the infrared illuminating wavelength), which is a key characteristic for their potential use in nanophotonics.

The in-plane anisotropic propagation of PhPs in $\alpha$-V$_2$O$_5$ can be more clearly observed by plotting the dispersions $\omega(k_i)$ ($i = x, y,$ corresponding to [100] and [001] crystal directions) extracted from monochromatic s-SNOM images (Fig. 2d). Although they look very similar, there is a clear separation between them (that is, for the same frequency $\omega_0$, we measured different wavevectors $k_i$), verifying that PhPs in RB$_1$ propagate with in-plane anisotropy. By plotting the complex-valued wavevector of the PhPs (Supplementary Information), we find that their phase velocity, $v_{p,i} = \omega_0/k_i$, is negative along both directions, which is indicated by negative $k_i$ values. Also, the slopes of the dispersion curves (Supplementary Information) yield very small group velocities ($v_{g,i} = (\partial k_i/\partial \omega)^{-1}$) of about $0.0009c$ in both directions (at $\omega_0 = 1026$ cm$^{-1}$, see Sup. Info), which is important for applications involving light-matter interactions[20].

To better describe the polaritonic response of $\alpha$-V$_2$O$_5$, we performed nanoscale Fourier transform infrared spectroscopy[2,21] (nanoFTIR) experiments by recording spectroscopic line scans (Methods) along the [100] and [001] in-plane crystal directions (Fig. 3a, left and right panels, respectively). We observe three spectral bands (RB$_{1-3}$) exhibiting a series of signal maxima within band limits corresponding to LO and TO phonon frequencies of $\alpha$-V$_2$O$_5$ (RB$_{1-3}$ in Fig. 1c, Supplementary Information) unveiling the existence of PhPs. In RB$_1$ we find that *i*) the signal maxima show a different spacing (corresponding to $\lambda_p$) along both [100] and [001] directions, and that *ii*) $\lambda_p$ increases with the frequency. These observations further reveal the in-plane anisotropic (elliptic) propagation and negative phase velocity of PhPs in the RB$_1$ of $\alpha$-V$_2$O$_5$. In contrast, in RB$_2$ we find that *i*) the signal maxima (dashed line in the figure) are only present along the [100] direction (there are no visible fringes along the [001] direction) and that *ii*) $\lambda_p$ decreases with the frequency. These observations indicate the excitation of PhPs with an in-plane hyperbolic propagation and a positive $v_p$. Finally, in RB$_3$ we find fringes, indicated by signal maxima (dashed lines in the figure), only along the [001] crystal direction (there are no visible fringes along the [100] direction), which points out to a similar in-plane hyperbolic behavior to that in RB$_2$ but along the orthogonal direction. Overall, these nanoFTIR results confirm that $\alpha$-V$_2$O$_5$ supports PhPs with in-plane anisotropic propagation (elliptic and hyperbolic along both orthogonal directions), thus adding another ~~a new~~ member to the library of vdW materials supporting PhPs[2,4,5].

As a metal oxide, α-$V_2O_5$, offers the possibility of being efficiently intercalated with alkali ions[22], which entails attractive prospects for energy technologies[23]. Intercalation has been recently demonstrated to be a promising low-temperature synthesis strategy to tune the physical and chemical properties of vdW materials such as $MoS_2$[24], $Bi_2Se_3$[25], or black phosphorus[26]. However, the effect of intercalation on the polaritonic response of a layered crystal has been barely studied, being a switching capability (originated by inducing a minimal polariton propagation in the intercalated area) together with a RB shift of a few wavenumbers the only polaritonic effects reported so far[4]. In the following, we study the effects of intercalation (using Na atoms by single crystal growth (Methods), as a difference to previous works based on aqueous solutions[4]) on the polaritonic response of α-$V_2O_5$. Figure 3b shows nanoFTIR line scans along the [100] and [001] in-plane directions (left and right panels, respectively) of an intercalated α'-(Na)$V_2O_5$ flake. ~~Excitingly~~ Interestingly, we observe a spectral band, RB'$_1$, that appears clearly red-shifted with respect to RB$_1$ in α-$V_2O_5$, and that shows periodic fringes of signal maxima indicating the existence of PhPs. These periodic signal maxima show a different spacing along the [100] and [001] directions, which increases with the frequency, revealing PhPs with in-plane anisotropic propagation (elliptic) and negative phase velocity in α'-(Na)$V_2O_5$. They are thus similar to PhPs in RB$_1$ of pristine α-$V_2O_5$ but shifted to smaller frequencies. To better analyze this effect we plot in Figure 4a the dispersions extracted from several monochromatic s-SNOM images for both pristine and intercalated flakes (Methods). We clearly observe that the dispersion of PhPs in RB`$_1$ (represented by dark green lines) of intercalated α'-(Na)$V_2O_5$ is strongly red-shifted (about 30 cm$^{-1}$ from center to center of both RBs), in comparison to the dispersion of PhPs in RB$_1$ of α-$V_2O_5$ (represented by bright green lines). This finding unambiguously demonstrates that a broad spectral shift of a RB can be originated by the intercalation of Na atoms into α-$V_2O_5$, thus demonstrating a key property for PhPs: spectral tunability via intercalation of the host material.

We note that the nanoFTIR image in Fig. 3b does not show any indication of other RBs that could be associated to RB$_2$ or RB$_3$ of α-$V_2O_5$. To better understand the polaritonic effects induced by the intercalation of Na atoms into α-$V_2O_5$ and the spectral shift obtained, we calculate from first principles (Supplementary Information) the phonon dispersions of α-$V_2O_5$, and α'-(Na)$V_2O_5$ crystals. The crystalline structures used in the calculations are illustrated in the top row of Figure 3, where the typical orthorhombic structure of α-$V_2O_5$ (left) is modified by the intercalation of Na atoms preferentially located in between the vdW layers (right). The extracted phonon modes (Supplementary Information) are then used in a Lorentz oscillators model to retrieve the theoretical permittivities for both crystalline structures, which are plotted in Figure 4b. In the plotted range, α-$V_2O_5$ shows three RBs (RB$_{1-3}$ in the figure) in qualitative agreement with the experimental permittivity shown in Fig. 1c, thus validating our *ab initio* calculations. For α'-(Na)$V_2O_5$ we also obtain three RBs (RB'$_{1-3}$), yet narrower and centered at shifted frequencies with respect to RB$_{1-3}$, respectively. Particularly, RB'$_1$ appears red-shifted for about 50 cm$^{-1}$ (from center to center) in good agreement with our results shown in the nanoFTIR images in Fig. 3b. On the other hand, mainly due to a modification of the oxygen effective charges along the [100] direction, RB'$_2$ appears blue-shifted and strongly narrower to the point of being almost imperceptible, which explains its absence in our experiments. Finally, RB'$_3$ is also narrower, being the LO phonon frequency strongly red-shifted, which also explains its absence in our measurements shown in Fig. 3b, as it lays out of our nanoFTIR spectral range.

Apart from the spectral shifts induced in the RBs (and thus in the PhPs dispersions) discussed above, the intercalation of Na atoms into the $\alpha$-V$_2$O$_5$ crystal lattice might have an influence on the PhPs anisotropic propagation and, more importantly, on their lifetimes. To compare the propagation of PhPs in $\alpha$-V$_2$O$_5$ and $\alpha$'-(Na)V$_2$O$_5$, we fabricated a metal antenna (gold disk) on top of both crystals and imaged their polaritonic activity by s-SNOM at frequencies within RB$_1$ and RB'$_1$, respectively (Figs. 5a, b). Due to its circular geometry, the gold disk can act as an efficient launcher of PhPs along all directions in the plane, thus enabling a direct investigation of the PhPs propagation. In both crystals, we observe fringes along all directions in the plane corresponding to the excitation of PhPs[27]. Interestingly, they both show a slightly longer $\lambda_p$ along the [001] direction, revealing a similar in-plane elliptic propagation of PhPs in both crystals. From this, we can draw the conclusion that intercalation has a negligible effect on the in-plane anisotropic propagation of PhPs in $\alpha$-V$_2$O$_5$.

Finally, we study and compare the PhPs lifetimes in both pristine $\alpha$-V$_2$O$_5$ and intercalated $\alpha$'-(Na)V$_2$O$_5$ crystals, which is crucial for validating intercalation as an effective material synthesis strategy for nanophotonics. To do this, we extracted s-SNOM line profiles (red crosses in Figs. 5c, d, taken at $\omega_0$ = 1010 cm$^{-1}$ and $\omega_0$ = 973 cm$^{-1}$, respectively) along the [001] direction of the $\alpha$-V$_2$O$_5$, and $\alpha$'-(Na)V$_2$O$_5$ flakes with similar thicknesses of 130 and 107 nm, shown in Figs. 5a, and b, respectively. By fitting them with an exponentially decaying sine-wave function corrected by the geometrical spreading factor[2,28] $\sqrt{x}$ (Supplementary Information) we extract decay lengths L$_{p[001]}$ of 1.40 μm for $\alpha$-V$_2$O$_5$ PhPs (yielding a propagation figure of merit[3] $Q$=Re($k$)/Im($k$) of 3.5) and 1.15 μm for $\alpha$'-(Na)V$_2$O$_5$ PhPs (yielding $Q$ = 2.5). With these values we obtain the lifetimes according to $\tau_{[001]}$ = L$_{[001]}$ /$v_g$, where the group velocities $v_g$ are taken from the PhPs dispersions (Fig. S4 in the Sup. Info.). We obtain $\tau_{[001]}$ = 6 ± 1 ps for PhPs in $\alpha$-V$_2$O$_5$, and $\tau_{[001]}$ = 4 ± 1 ps for PhPs in intercalated $\alpha$'-(Na)V$_2$O$_5$. These notable values of the lifetimes (obtained despite a short L$_p$ due to the ultra-slow PhPs group velocity of 0.0007c) reveal the low-loss nature of PhPs in $\alpha$-V$_2$O$_5$, which can be related to the fact that $\alpha$-V$_2$O$_5$ is close to be isotopically pure, since natural abundance oxygen and vanadium have both isotopic purities of 99.7%, and, more importantly, that the intercalation process followed in this work not only allows for a large spectral shift of PhPs, but also for preserving their low-loss nature (despite a significant reduction, the lifetime is still in the range of ps).

In conclusion, this work demonstrates intercalation of atoms (Na) in a vdW crystal ($\alpha$-$V_2O_5$) as an efficient technological approach to achieve a broad spectral shift of PhPs with ultra-long lifetimes. Considering that a large variety of ions and ion contents can be intercalated in layered materials[29], we envision on-demand spectral response of PhPs in vdW materials, eventually allowing for covering the whole mid-IR range, critical for the emerging field of phonon polariton photonics. We also note that $\alpha$-$V_2O_5$, as a semiconductor, can be electrically doped to support plasmon polaritons, which, via coupling to LO phonons[30], could eventually allow for a dynamic tuning of PhPs spectrally shifted by intercalation.

**METHODS**

**s-SNOM**

We used a scattering type scanning near-field optical microscope (s-SNOM, from Neaspec GmbH) for infrared nanoimaging. This system employs a Pt-Ir coated atomic force microscope (AFM) tip illuminated with infrared light as a launcher and recorder of polaritons, yielding simultaneously near-field images and topography. The metallized AFM tip is illuminated with p-polarized light of frequency $\omega$ from a quantum cascade laser. The tip oscillates vertically at a frequency $\Omega \approx 270$ kHz with an amplitude of about 100 nm. The tip-launched polaritons reflect at the $V_2O_5$ flake edges and produce polariton standing wave interferences, which are imaged by recording the light scattered by the tip. This is carried out with a pseudo-heterodyne interferometer that demodulates the detector signal at high harmonics $n\Omega$ (normally 3rd harmonics named as $s_3$) and provides background-free detection. The metal coated AFM tip provides wavelength independent resolution.

**nanoFTIR**

For the nanoFTIR images, we employed the nanoFTIR module for the s-SNOM system. We used an Au coated AFM tip illuminated by a super-continuum laser and the scattered light was recorded with an asymmetric Fourier transform spectrometer. By recording point spectra as a function of the tip position, we obtained high-resolution spectral line scans.

**$\alpha$-$V_2O_5$ and $\alpha$'-(Na)$V_2O_5$ sample growth and preparation**

Orthorhombic layered $\alpha$-$V_2O_5$ with Van der Waals structure was synthesized via a purification method and subsequent single crystal growth, as described in references [14, and 31]. Sodium-intercalated $\alpha$'-(Na)$V_2O_5$ single crystals were grown in a similar way as described in reference [32]. Briefly, a mixture of $NaVO_3$ (1.843g) and $VO_2$ (0.1564g) powder was put into a Pt crucible with a lid and sealed under vacuum in a quartz tube. The mixture was heated to 800 ºC at a rate of 100 ºC/h and kept at 800 ºC for 0.5 h. It was then cooled to 740 ºC at a rate of 1 ºC/h, kept at 740 ºC for 3 days, slowly cooled to 600 ºC at a rate of 1.5 ºC/h, and finally cooled to room temperature at a rate of 300 ºC/h to yield $\alpha$'-(Na)$V_2O_5$ crystals. Temperature dependent magnetic susceptibility measurement of the as-grown α'-(Na)V₂O₅ crystal verified a spin-Peierls transition at ~35K, in agreement with the literature[32] and indicating adequately intercalated α'-(Na)V₂O₅. Bulk α-V₂O₅ and α'-(Na)V₂O₅ crystals were thinned down by mechanical exfoliation using Nitto blue tape leading to thin slabs with thicknesses in the range from 100 nm to a few micrometers, which are transferred on top of a 300-nm-thick SiO₂ layer deposited on a Si substrate. For maximum yield of thin devices, substrates were heated to 90 ºC for 10 minutes during transfer.

**Dielectric function of α-V₂O₅**

The principal values of the diagonal permittivity tensor, $\varepsilon_x$, $\varepsilon_y$ and $\varepsilon_z$ are approximated with a three-parameter Drude–Lorentz permittivity:

$$\varepsilon_{a(\omega)} = \varepsilon_{a,\infty} \frac{(\omega_{LO}^a)^2 - \omega^2 + i\gamma^a \omega_{LO}^a \omega}{(\omega_{TO}^a)^2 - \omega^2 + i\gamma^a \omega_{TO}^a \omega},$$

where $a = x, y, z$, $\omega_{TO}$ and $\omega_{LO}$ refer to the transversal (TO) and longitudinal (LO) phonon frequencies, respectively, $\gamma$ denotes the damping constant and $\varepsilon_{a,\infty}$ is the high frequency permittivity. The values of the constants are: $\omega_{TO}^x$= 765 cm⁻¹, $\omega_{LO}^x$= 952 cm⁻¹, $\gamma^x$= 40 cm⁻¹, $\varepsilon_{x\infty}$= 6.6; $\omega_{TO}^y$= 506 cm⁻¹, $\omega_{LO}^y$= 842 cm⁻¹, $\gamma^y$= 19 cm⁻¹, $\varepsilon_{y\infty}$= 6.1; $\omega_{TO}^z$= 976 cm⁻¹, $\omega_{LO}^z$= 1037 cm⁻¹, $\gamma^z$= 2.0/1.5 cm⁻¹, $\varepsilon_{z\infty}$= 3.9.

The values of $\omega_{LO}^a$ and $\omega_{TO}^a$ were adjusted by fitting FTIR and s-SNOM measurements with transfer-matrix calculations, while $\varepsilon_{a,\infty}$ were obtained from *ab initio* calculations (Supplementary Information). The values of $\gamma^a$ were taken from ref. [15]. $\gamma^z$ = 1.5 cm⁻¹ is also considered (Supplementary Information) to better fit the experimental lifetime values and in analogy to ref [33].

**Structural Characterization**

α-V₂O₅ and α'-(Na)V₂O₅ local structures were analyzed by High Resolution Transmission Electron Microscopy (HRTEM) using a Tecnai F20 ST (FEI) operating at 200 kV. X-ray diffraction was performed using a Bruker Discover D8 with VÅNTEC-500 solid state detector, using a Co K-alpha X-ray source with wavelength 1.788 nm.

**Methods References**

## DATA AVAILABILITY

The data that support the findings of this study are available from the corresponding authors on reasonable request.

## CODE AVAILABILITY

The code employed in this work to perform all calculations is available from the corresponding authors on reasonable request.

**ACKNOWLEDGEMENTS**

J.T-G and G.Á-P acknowledge support through the Severo Ochoa Program from the Goverment of the Principality of Asturias (PA-18-PF-BP17-126 and PA-20-PF-BP19-053, respectively). J.M-S. acknowledges support through the Clarín Programme from the Government of the Principality of Asturias and a Marie Curie-COFUND grant (PA-18-ACB17-29). K.C., X.P.A.G., H.V., and M.H.B. acknowledge the Air Force Office of Scientific Research (AFOSR) Grant FA 9550-18-1-0030 for funding support. I.E. acknowledges financial support from the Spanish Ministry of Economy and Competitiveness (FIS2016-76617-P). A.Y.N. acknowledges the Spanish Ministry of Science, Innovation and Universities (national project MAT2017-88358-C3-3-R) and Basque Government (grant No. IT1164-19). Q. B. acknowledges the support from Australian Research Council (ARC, FT150100450, IH150100006 and CE170100039). P.A-G. acknowledges support from the European Research Council under Starting Grant 715496, 2DNANOPTICA.

**AUTHOR CONTRIBUTIONS**

P.A-G. and J.T-G. conceived the initial measurements on $\alpha$-$V_2O_5$. P.A-G. and J.M-S. supervised the project. J.T-G, and J.D carried out the near-field imaging experiments with the help of M.A., S.L. and W.M. G.A-P., A.B., R.H., Q.B. and A.Y.N. participated in data analysis. P. A-G wrote the manuscript with input from J.T-G, G.A-P, J.M-S, A.Y.N, J.D, Q.B, I.E., X.P.A.G, K.C. and R.H. I.E carried out the *ab initio* calculations. K.K., T.K., K.C., and X.P.A.G. contributed to material synthesis and sample preparation. H.V. and M.H. B. performed the TEM characterization. K.C. performed XRD indexing and characterization. I.P. contributed to sample fabrication.

**COMPETING FINANCIAL INTERESTS**

R.H. is co-founder of Neaspec GmbH, a company producing scattering-type scanning near-field optical microscope systems, such as the one used in this study. The remaining authors declare no competing interests.

**FIGURES**

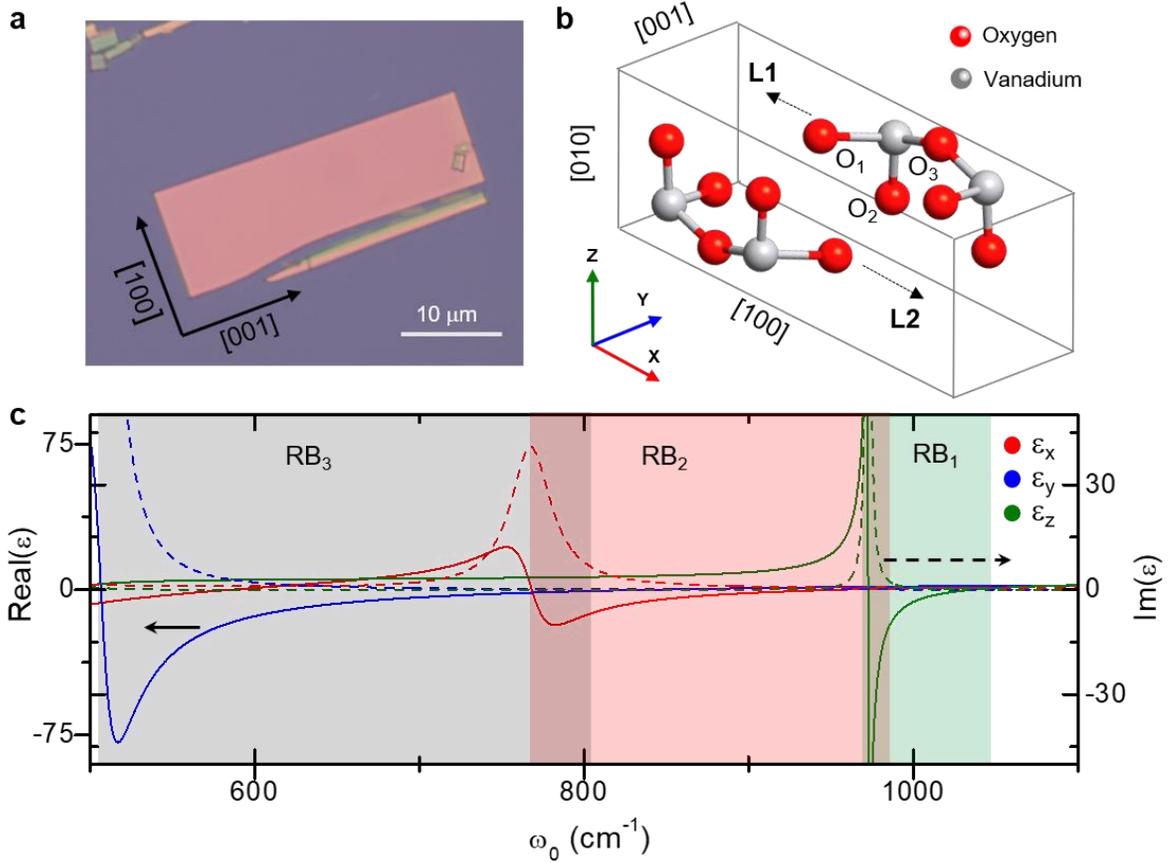

**Figure 1**: **Physical properties of α-V$_2$O$_5$. a.** Optical image of an α-V$_2$O$_5$ flake. α-V$_2$O$_5$ crystals show a rectangular shape owing to their anisotropic crystal structure. Labelled arrows indicate the [100] and [001] crystal directions. **b.** Schematic of the crystalline unit cell of α-V$_2$O$_5$; the lattice constants along the principal x, y, and z axes are $a$ = 1.15 nm, $b$ = 0.36 nm and $c$ = 0.44 nm, respectively. Grey spheres represent the vanadium atoms and red spheres represent the oxygen atoms. The three oxygen positions of asymmetric V–O bonds along different crystalline axes are denoted O$_{1-3}$. L1 and L2 indicate the two vdW layers weakly bound along the [010] crystal direction. **c.** Real (continuous lines) and imaginary part (dashed lines) of the permittivity (see Methods) along the principal x ([100]), y ([001]), and z ([010]) axes (red, blue, and green lines, respectively). The Reststrahlen bands RB$_1$, RB$_2$, and RB$_3$ are indicated in green, red and grey, respectively.

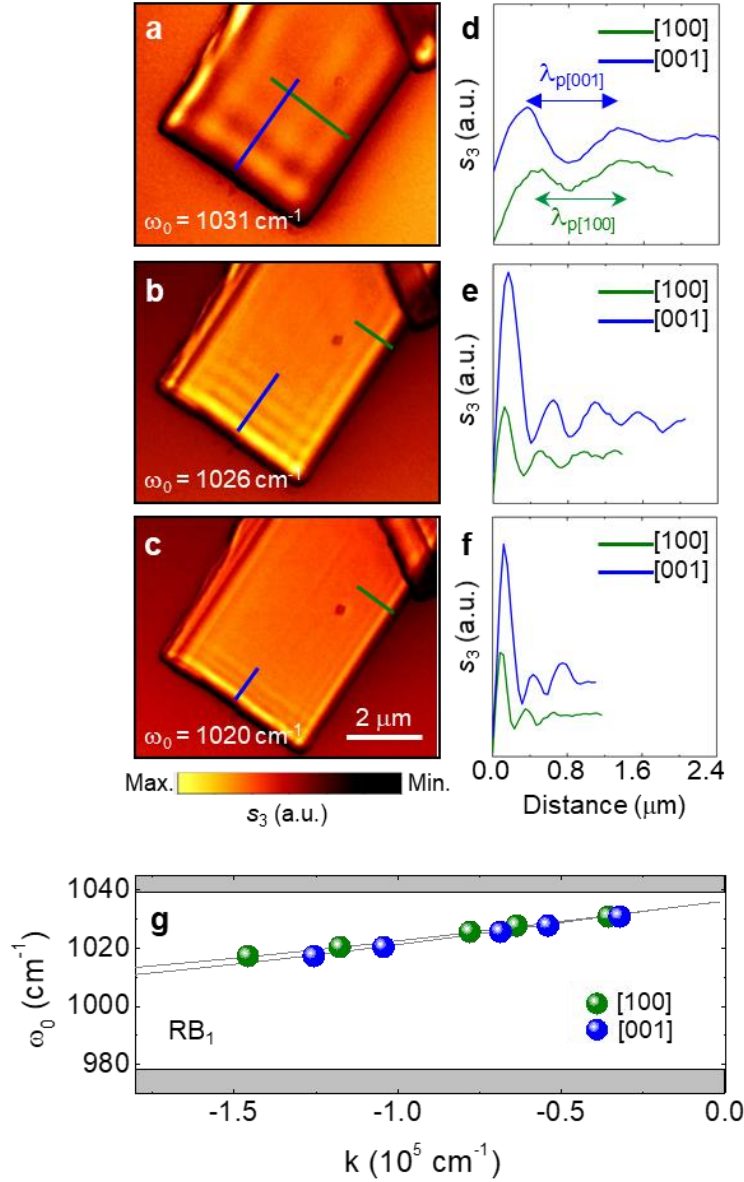

**Figure 2: Real-space imaging of a α-V$_2$O$_5$ flake. a-c.** Near-field amplitude images s$_3$ of an α-V$_2$O$_5$ flake with thickness $d$ = 105 nm at incident frequencies ω$_0$ = 1031 cm$^{-1}$, 1026 cm$^{-1}$, and 1020 cm$^{-1}$, respectively. **d-f** Profiles along the [100] (green lines) and [001] (blue lines) directions, extracted from the near-field amplitude images in **a-c**, respectively.

$\lambda_{p[100]}$ and $\lambda_{p[001]}$ are the polariton wavelength along the [100] and [001] direction, respectively. **g.** Dispersion of PhPs along the [100] (green symbols) and [001] (blue symbols) directions in the RB$_1$. Grey lines are guides to the eye. Grey shaded areas indicate the spectral regions outside the RB.

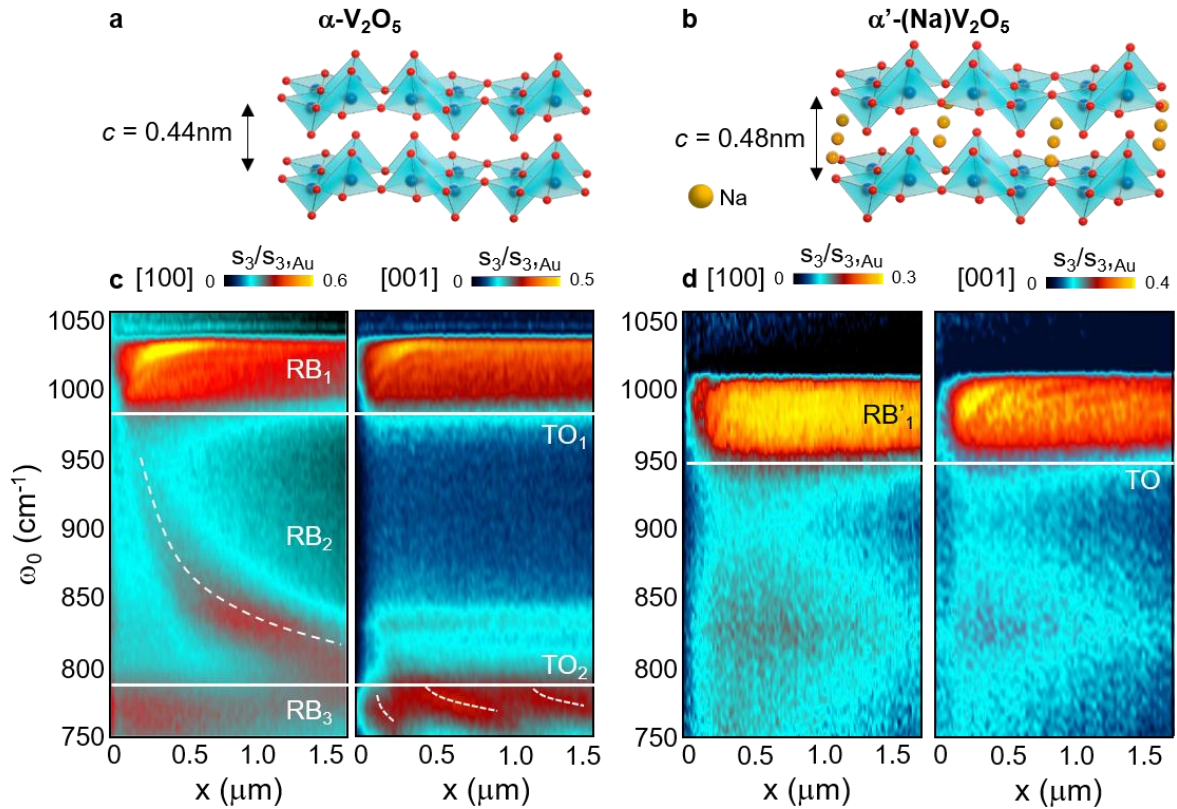

**Figure 3: Real-space nano-spectroscopy of $\alpha$-V$_2$O$_5$ and intercalated $\alpha'$-(Na)V$_2$O$_5$ flakes**. **a.** Illustration of the orthorhombic lattice structure of layered $\alpha$-V$_2$O$_5$ (red spheres, oxygen atoms; blue atoms, vanadium atoms; blue pyramids, polyhedral defined by the oxygen atoms). The orthorhombic structure is based on bilayers of distorted VO$_5$ square pyramids stacked along the [010] direction - with interlayer distance $c$ = 0.44 nm - via vdW interactions. **b.** nano-FTIR spectral line scans along the [100] and [001] directions of a $\alpha$-V$_2$O$_5$ flake showing the near-field amplitude $s_3/s_{3,Au}$ (normalized to the near-field amplitude on Au, $s_{3,Au}$) as a function of distance between tip and flake edge. Solid horizontal lines mark the approximate transversal phonon modes in $\alpha$-V$_2$O$_5$ (TO$_1$, 975 cm$^{-1}$; TO$_2$, 770 cm$^{-1}$), separating RB$_{1-3}$. Dashed lines are guides-to-the-eye of signal maxima. The flake thickness is $d$ = 245 nm **c.** Illustration of the orthorhombic lattice structure of layered $\alpha'$-(Na)V$_2$O$_5$ (red spheres, oxygen atoms; blue atoms, vanadium atoms; yellow atoms, sodium atoms; blue pyramids, polyhedral defined by the oxygen

atoms). The orthorhombic structure is based on bilayers of distorted VO$_5$ square pyramids with sodium atoms intercalated and stacked along the [010] direction - with interlayer distance $c$ = 0.48 nm - via vdW interactions. **d.** nano-FTIR spectral line scans along the [100] and [001] directions of a α'-(Na)V$_2$O$_5$ flake showing the near-field amplitude $s_3$ (normalized to the near-field amplitude on Au, $s_{3,Au}$) as a function of distance between tip and flake edge. Solid horizontal lines mark the approximate transversal phonon modes in α'-(Na)V$_2$O$_5$ (TO, 950 cm$^{-1}$), defining RB'$_1$. The flake thickness is $d$ = 150 nm. The scale in the colour bars of **b** and **d** is linear.

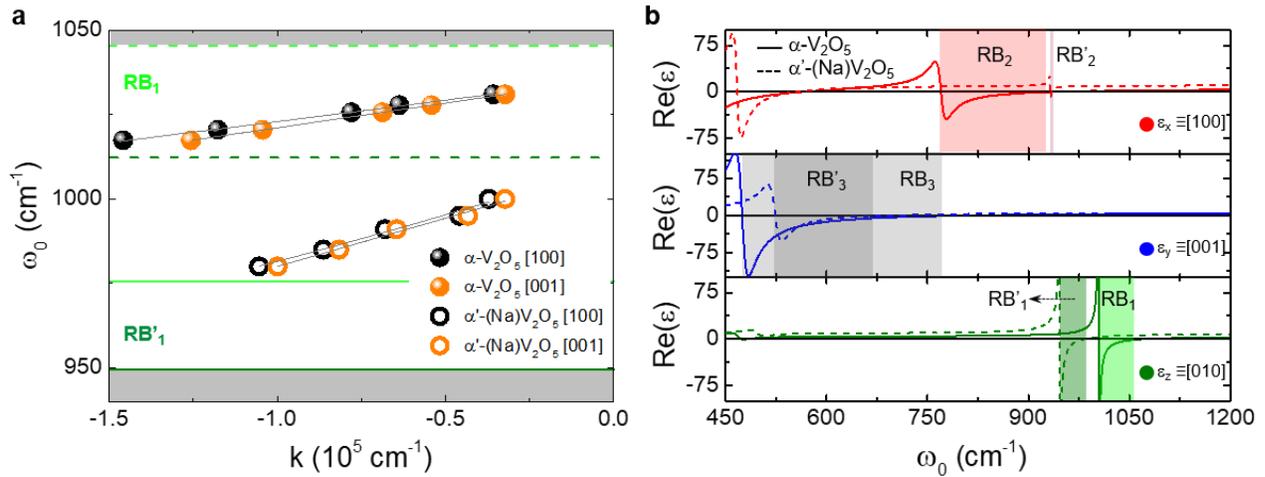

**Figure 4: PhPs dispersion and *ab initio* permittivity in α-V$_2$O$_5$ and intercalated α'-(Na)V$_2$O$_5$ crystals. a.** Dispersion of PhPs along the [100] and [001] directions in a α-V$_2$O$_5$ (full symbols) flake with thickness $d$ = 105 nm and a α'-(Na)V$_2$O$_5$ (empty symbols) flake with thickness $d$ = 190 nm. Dashed and continuous horizontal lines mark the approximate TO and LO phonon modes in α-V$_2$O$_5$ (TO, 980 cm$^{-1}$; LO, 1040 cm$^{-1}$), and α'-(Na)V$_2$O$_5$ (TO, 945 cm$^{-1}$; LO, 1015 cm$^{-1}$), respectively. Grey lines are guides to the eye. Grey shaded areas indicate the spectral regions outside the RBs. **b.** Real-part of the permittivities for α-V$_2$O$_5$ (continuous lines) and α'-(Na)V$_2$O$_5$ (dashed lines) extracted from *ab initio* calculations along the principal x, y, and z axes (red, blue, and green lines, respectively). The Reststrahlen bands RB$_{1-3}$, and RB'$_{1-3}$ for α-V$_2$O$_5$ and α'-(Na)V$_2$O$_5$, are indicated in bright and dark color, respectively. Green shaded regions represent RB$_1$ and RB'$_1$; Red shaded regions represent RB$_2$ and RB'$_2$ and grey shaded regions represent RB$_3$ and RB'$_3$.

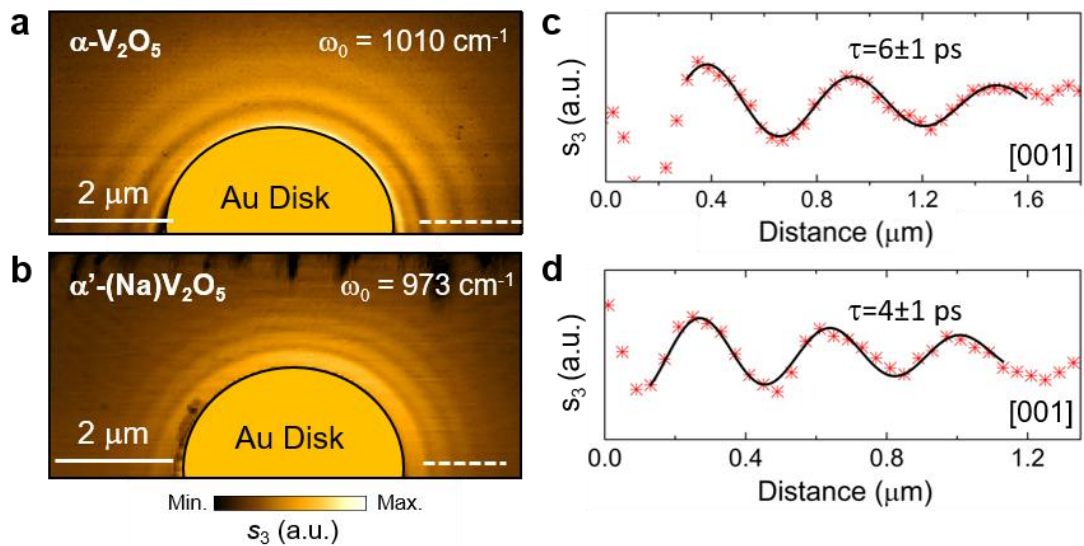

**Figure 5: Anisotropy and lifetimes of PhPs in $\alpha$-V$_2$O$_5$ and intercalated $\alpha$'-(Na)V$_2$O$_5$ flakes. a,b.** Near-field amplitude images s$_3$ of $\alpha$-V$_2$O$_5$ and $\alpha$'-(Na)V$_2$O$_5$ flakes with thicknesses $d$ = 130 nm, and $d$ = 107 nm at illuminating frequencies $\omega_0$ = 1010 cm$^{-1}$ (RB$_1$) and $\omega_0$ = 973 cm$^{-1}$ (RB'$_1$), respectively. A gold disk (half of it shown in the image for convenience) is used as an antenna for efficient launching of PhPs along all in-plane directions. **c,d.** s-SNOM line traces (showing the amplitude s-SNOM signal σ$_3$; Methods) along the [001] direction of $\alpha$-V$_2$O$_5$ and $\alpha$'-(Na)V$_2$O$_5$ flakes in a and b (white dashed lines). Damped sine-wave functions (black solid lines) were fitted to the data (Supplementary Information). Lifetimes $\tau$ = 6 ± 1 ps and $\tau$ = 4 ± 1 ps are obtained, respectively.

15